\documentclass[a4paper, USEnglish]{article}

\usepackage{microtype}
\usepackage{todonotes}

\usepackage{color} 

\graphicspath{ {images/} }
\usepackage{graphicx}
\usepackage{caption}

\usepackage{amsmath}
\usepackage{amsfonts}

\usepackage{listings}
\usepackage{array}
\usepackage{adjustbox}
\usepackage{subfloat}
\usepackage{hyperref}
\usepackage[affil-it]{authblk}

\renewcommand{\todo}[1]{}
\usepackage{xspace}
\newcommand{\tIFC}{\ensuremath{t_\mathit{IFC}}\xspace}

\title{IIFA: Modular Inter-app Intent Information Flow Analysis of Android Applications}

\definecolor{dkgreen}{rgb}{0,0.6,0}
\definecolor{gray}{rgb}{0.5,0.5,0.5}
\definecolor{mauve}{rgb}{0.58,0,0.82}
\begin{document}

\author{Abhishek Tiwari, Sascha Gro\ss, Christian Hammer \\ {\{tiwari, saschagross, chrhammer\}@uni-potsdam.de}}
\affil{University of Potsdam, Germany}

\maketitle

\begin{abstract}
Android apps cooperate through message passing via intents. However, when apps do not have identical sets of privileges inter-app communication (IAC) can accidentally or maliciously be misused, e.g., to leak sensitive information contrary to users’ expectations. Recent research considered static program analysis to detect dangerous data leaks due to inter-component communication (ICC) or IAC, but suffers from shortcomings with respect to precision, soundness, and 
scalability.

To solve these issues we propose a novel approach for static ICC/IAC analysis. 
We perform a fixed-point iteration of ICC/IAC summary information to precisely resolve intent communication with more than two apps involved. We integrate these results with information flows generated by a baseline (i.e.~not considering intents) information flow analysis, and resolve if sensitive data is flowing (transitively) through components/apps in order to be ultimately leaked. Our main contribution is the first fully automatic sound and precise ICC/IAC information flow analysis 
that is scalable 
for realistic apps due to modularity, avoiding combinatorial explosion: Our approach determines communicating apps using short summaries rather than 
inlining intent calls, which often requires simultaneously analyzing all tuples of apps.

We evaluated our tool IIFA in terms of scalability, precision, and recall. Using benchmarks we establish that precision and recall of our algorithm are 
considerably better than prominent state-of-the-art analyses for IAC. But foremost, applied to the 90 most popular applications from the Google Playstore, IIFA demonstrated its scalability to a large corpus of real-world apps. IIFA reports 62 problematic  ICC-/IAC-related information flows via two or more apps/components.

\end{abstract}

\section{Introduction}


Mobile devices are an attractive target for all kinds of dubious activities as they store a plenitude of sensitive data. 
Therefore, protecting the information stored on smartphones from unauthorized access has become imperative. Manufacturers implemented a permission system that lets the user decide which privileges an app may have, e.g., to access sensitive information or to communicate via certain channels. However, permissions cannot restrict information flow once access to sensitive information or a communication channel has been granted.
%
Given that the number of available Android apps is enormous 
and that market places are 
lacking thorough security checks before publishing an app\footnote{Even Google's official security analysis has been circumvented~\cite{oberheide2012dissecting}},
several cases have been reported where vulnerable or malicious apps leak sensitive information~\cite{zhou2012dissecting,wei2014amandroid}.

To protect sensitive information on Android, various information flow control (IFC) analyses have been developed. These analyze the (potential) flow of information in apps and report a warning if a flow from a sensitive data source to an untrusted/public data sink (like sending sensitive information to the internet) is determined. Information flow is not restricted to a single component, but occurs frequently between components of the same~\cite{li2015iccta,gordon2015information} and even different apps~\cite{wei2014amandroid}. Our study using the top 90 apps from the Google play store revealed more than 10,000 inter-component calls. Scrutinizing the flows between components therefore becomes imperative.

Android's ICC mainly leverages so-called \emph{intents}. The major challenge in identifying IFC through intents is identifying which information flows from one component to another. Leveraging static analysis is non-trivial because the receiver and the intent data may be unknown at analysis time, being strings that might be composed at runtime. 

Some tools consider intents during information flow analysis~\cite{li2015iccta,wei2014amandroid} but suffer from multiple shortcomings:
Both approaches basically inline a synthetic ``main'' method that models the lifecycle of the receiving component/app into the sender of the intent. However, in order to match senders and receivers they merely verify that the intent action (or similar receiver-identifying data) matches. Neither of those two approaches actually determines whether the receiver and the sender use the same type or key in the key-value communication scheme of \emph{extra data} transmitted via intents. Thus, either some or even all receivers in the inlined lifecycle might not be eligible to read the transmitted data, which can thus result in many spuriously reported data leaks. A more accurate matching could probably be added to these approaches, but only if there is merely one sender and one receiver statement in any given pair of communicating components. In case of multiple sender and/or receiver statements one matching pair might still induce a quadratic number of spurious flows.

Inlining several sender and receiver apps into one huge app to analyze (e.g.~90 apps like in our evaluation study) requires a very precise analysis in order not to magnify the imprecision described in the last paragraph 
(e.g., context-sensitivity to match multiple senders to the same component lifecycle or even that one sender is only problematic if it has received sensitive intent data from another app). Besides, inlining several apps into one raises scalability issues as even single app ICC is challenging in terms of scalability~\cite{li2015iccta}. Therefore the straight forward would be to eagerly analyze all pairs of apps. Note that at least a quadratic number of combinations must be analyzed to include the effects of IAC to IFC. Realistically, intent communication can involve more than two apps, further aggravating the combinatorial explosion of merging-based approaches. Besides, merging itself is impractical in two dimensions: Merging APKs or other internal data structures does not scale to realistic apps in our experience, and even if it does, the complexity to analyze the merged app inflates, but as most combinations of apps do not communicate via intents the whole effort is mostly futile. Simultaneously the merging process itself may introduce spurious data flow paths, increasing analysis imprecision.

Finally, related work suffers from requiring access to source code (or even source code annotations)~\cite{appscan,huang2016detecting,barros2015static,octeau2015composite}, lacking support for string analysis~\cite{arzt2014flowdroid,li2015iccta}, and lacking support for certain sink functions~\cite{arzt2014flowdroid,li2015iccta,appscan,klieber2014android}. All of these drawbacks lead to insufficient precision and soundness issues when these analyses are applied to real-world apps. 

\textbf{Our Contributions.}
In this work we propose a novel information-flow analysis that evades combinatorial explosion of potential communication partners, while precisely matching type and key information of intent data. Our approach can predict which combinations of apps communicate by separating analysis of apps and matching of communication partners. 
In a first step 
we create a database of summary information about senders of intents, their characteristics including types and keys, and outbound intent data, as well as apps registered to receive certain implicit intents. This information can then be matched in a subsequent step to identify potential communication partners. Further we propose a novel matching algorithm, based on a baseline IFC analysis providing potential intra-app flows (including a program slices of the receiver's key value) for all potential intent receivers. We use senders' outbound intent data as input to the information flows identified in respective receivers, which eliminates the need for inlining or merging apps and thus combinatorial explosion, as only summaries of actual communication partners are subsumed.
In case multiple apps are involved in intent communication our approach performs a light-weight fixed point iteration through the DB information.
Note that our tool is not a stand-alone IFC analysis tool. Rather, IIFA leverages flows and slices generated by other IFC analyzers. As these tools are already heavily engineered for the intra-app case, we concentrated on the peculiarities of intent communication and evasion of inlining and combinatorial explosion. 

As a noteworthy novelty, our approach is modular and thus compositional with respect to app installation. Whenever a new (version of an) app is available for analysis, the database is updated (in case of new version) or extended (new app) to include the intents broadcast or received by this app. Only the new app has to be (re-)analyzed, as well as combinations with flows identified in potential receivers.
We compute precise communication paths between components handling complex control flows such as in callback methods (see section~\ref{sec:callback}). 
As intent targets are specified via a string parameter, our approach can resolve common string manipulations
, which improves our precision significantly for regular (non-obfuscated) apps.
As a minor contribution we took great effort to handle the full spectrum of intent communication, supporting explicit and implicit intents as well as dynamically created intent receivers. All previous approaches miss at least one of these features, leading to unsoundness and imprecision. Our analysis is fully automated and does not require the source code of the app under analysis. We aim to answer the following research questions:
\begin{itemize}
	\item \textbf{RQ1:} \textit{How are the precision and soundness of our approach with respect to the state of the art analyses?}
	\item \textbf{RQ2:} \textit{Does our approach scale to a realistic corpus of real-world apps?}
	\item \textbf{RQ3:} \textit{How do common real-world apps communicate through IAC?}
\end{itemize}

We implemented our approach as a tool called \emph{IIFA} and evaluated it on DroidBench, the IccTA extension of DroidBench, ICC Bench, and a large set of apps from the Google Playstore. We compared our results with multiple related analysis tools. Our tool (combined with an external baseline intra-component IFC analysis)  achieves perfect precision and soundness on all benchmark sets, being more than on par with related IFC tools that consider intent communication. Additionally, we demonstrate the superior IAC precision with experiments and an evaluation 
applying IIFA to the 90 most downloaded Playstore apps\todo{check}. Further, our experiments demonstrate that due to its compositionality IIFA's execution time scales well even to large real-world apps. In summary, we provide the following contributions:
\begin{itemize}
\item \emph{Compositional DB-backed Analysis.} We propose a modular analysis approach for intent communication based on summaries containing intent senders, receivers, and the exact intent characteristics including types and keys of data transmission. To that end, we model all publicly known intent-based communication schemes precisely.
\item \emph{Novel Matching Algorithm.} We present a novel algorithm which matches intent senders with intent receivers based on the summaries, and subsumes transmitted data into the receiver's intra-component information flows to report potential dangerous inter-component and -app information flows. This matching requires no eager pairwise analysis but only investigates potential communication partners. Thus each app is only analyzed once by IIFA and potentially as well by an intra-app IFC analysis.

\item \emph{Evaluation of IIFA.} We implemented our analysis (IIFA) and evaluated it on multiple large-scale datasets. The evaluation shows that our analysis is on par or better than the most relevant previous work in terms of precision and recall with respect to previously presented benchmarks. We demonstrate our improved ICC precision with IAC benchmarks and a real-world evaluation analyzing the top 90 real-world apps.\todo{check} 
Finally we demonstrate that we can effectively evade combinatorial explosion analyzing these 
apps in less than two minutes per app (in addition to the traditional IFC analysis) and identifying 62 potentially dangerous information flows through ICC.
\item \emph{IAC study.} We performed a study regarding the use of ICC/IAC in Android and present our results outlining IAC patterns.
\end{itemize}


\section{Background}

\subsection{Android Components}

Android apps are written (mostly) in Java, but instead of defining a \emph{main} method they consist of four component types: 
\emph{Activities} are user interfaces to be interacted with. \emph{Services} run in the background, intended for computationally expensive operations. \emph{Broadcast receivers} register themselves to receive system or app events. \emph{Content providers} provide data via storage mechanisms.
Each app defines a manifest file (\emph{AndroidManifest.xml}) providing essential information about the app, e.g., 
components and their capabilities are defined in the manifest file. 

Apps are compiled to Dalvik bytecode~\cite{dalvikDocumentation}, which is specialized for execution on Android. Together with additional metadata and resources Dalvik bytecode is compressed into an Android Package (APK) that can be published in market places, such as the Google Playstore. Oberheide and Miller~\cite{oberheide2012dissecting} demonstrated that the security analysis on the Playstore can easily be circumvented. Even though Google's security mechanisms are constantly evolving, potential for malicious and vulnerable software in the Google Playstore remains.


\subsection{Android Intents}
Android provides a dedicated mechanism for two components to communicate. A component can send an \emph{intent} as a message, e.g., to notify another component of an event, trigger an action of another component, or transmit information to another component. Note the universal nature of intents on Android: Intents can be sent from the system to apps (and vice versa), from one app to another (inter-app communication, IAC), or even from one component to another within the same app (intra-app-communication, ICC)
~\cite{intentDocumentation}.

Additional information can be associated with an intent: 
The \emph{intent action} specifies an action supposed to be performed by the receiving component. A component can register to receive intents with a specific intent action by declaring an intent filter in the manifest file (see discussion of Listing~\ref{listing:intenttwo} below). The intent's sender is unknown on the receiver side. The \emph{target component} mandates a specific receiver for an intent. Setting \emph{intent extra data} adds additional information to be used as parameters by the receiving component.

It is important to note that none of this information is mandatory. When a target component is specified an intent is called \emph{explicit}, otherwise \emph{implicit}. 
Explicit intents are delivered to the given target component only, while implicit intents can be delivered to any component with a matching intent filter
. If multiple components could receive an implicit intent, the user is asked to resolve the intent manually, generally displaying a list of potential receiver apps. Li et al.~\cite{li2015iccta} found that at runtime 40.1\% of the intents in Google Playstore apps are explicit intents. 
Broadcast intents are relayed to every component registered for an intent action instead of only one of them. 
As intents are the universal means of inter-component communication their analysis becomes critical. In this work we propose a modular approach to precisely analyze information flow through Android intents.


\section{Motivation}
\label{example}
\begin{lstlisting}[caption=App A - Sender: OutFlowActivity,label={listing:intent}, float]
@\label{lst:line1_1}@TelephonyManager tel = (TelephonyManager) getSystemService(TELEPHONY_SERVICE);
@\label{lst:line1_2}@String imei = tel.getDeviceId(); // source
@\label{lst:line1_3}@Intent i = new Intent("CUSTOM_INTENT.ACTION");
@\label{lst:line1_4}@i.putExtra("data", imei);
@\label{lst:line1_5}@startActivity(i); // sink	
\end{lstlisting}
\begin{lstlisting}[caption=App B - Receiver: AndroidManifest.xml, label={listing:intenttwo}, float]
<intent-filter>
  <action android:name="CUSTOM_INTENT.ACTION"/>
  <category android:name = "android.intent.category.DEFAULT" />
</intent-filter>	
\end{lstlisting}
\begin{lstlisting}[caption=App B - Receiver: InFlowActivity, label={listing:intentthree}, float]
@\label{lst:line2_1}@Intent i = getIntent();
@\label{lst:line2_2}@String imei = i.getStringExtra("data");
@\label{lst:line2_3}@smsManager.sendTextMessage("1234567890", null, imei, null, null); // sink	
\end{lstlisting}

In this section we will describe an example workflow of intents. We will then discuss the inefficiency of the current state-of-the-art analysis tools. 
In Listing~\ref{listing:intent}, \emph{App A} initiates IAC in the \emph{OutFlowActivity} class. An implicit intent \emph{i} is created (line~\ref{lst:line1_3}) with the intent action \emph{``CUSTOM\_INTENT.ACTION"}. The \emph{putExtra} method (line~\ref{lst:line1_4}) associates additional data from the variable \emph{imei} with the key \emph{``data"} in this intent. The device id stored in variable \emph{imei} (lines~\ref{lst:line1_1} and~\ref{lst:line1_2}) is sensitive data, as the device can be uniquely identified by this number. The intent is finally triggered via a \emph{startActivity} call such that it can be received by registered receivers. On the receiver side the intent extra data can be extracted by the receiver and (ab-)used in any way permissible to that app.

Listing~\ref{listing:intenttwo} and~\ref{listing:intentthree} show the code snippets of an example receiver for the above intent. In the manifest file (Listing~\ref{listing:intenttwo}) the receiver declares its capability to support intent filter (\emph{``CUSTOM\_INTENT.ACTION"}). Listing~\ref{listing:intentthree} extracts the received intent data corresponding to the key ``data" via a \emph{getStringExtra} method call. This data flows to a data sink (line~\ref{lst:line2_3}) where it is being leaked off the device. Since the received data is a sensitive information with respect to App~A, analysis tools should report this as a potential data leak. Observe that these two apps must be related by an analysis in order to identify the leak and that the precision for determining intent data  crucially influences the precision to determine related apps and which data is transmitted. 

Current analysis tools for ICC~\cite{li2015iccta}/IAC~\cite{wei2014amandroid} only match the identifying string (like the intent action in Listing~\ref{listing:intent}-~\ref{listing:intentthree}) and ignore the key ``data'' or the type (in our case \emph{String}, see line~\ref{lst:line2_2} of the receiver), which must also match for data to be transmitted. If multiple receivers are present in one component then adding checks for these conditions is non-trivial as these approaches basically inline the receiver into the sender, thus the same intent would be used for all receivers even if the key or type of the \emph{getExtra}-method differs, resulting in spurious reported information flows.
Simply merging/inlining all apps installed on a device is prohibitively expensive, and may result in impossible flows created as a result of the merging process, thus in practice these approaches may resort to eagerly inlining pairs, triples, \dots\ of communicating apps, leading to combinatorial explosion of analysis targets. Furthermore, in a realistic scenario apps may be installed on a device at any time. Thus, whenever a new app (version) arrives, these tools need to join apps again. 

To overcome these limitations, our analysis is designed in a modular way: It remembers summaries of the intent characteristics (including key and type) of each app in a database and applies this knowledge to (intra-app) information flows determined in receiving apps. Due to the database our analysis can recursively resolve dependences when more than two apps are involved in intent communication. 

\section{Methodology}

\begin{figure}[tb]
\includegraphics[width=\linewidth,viewport=20 30 475 135,clip]{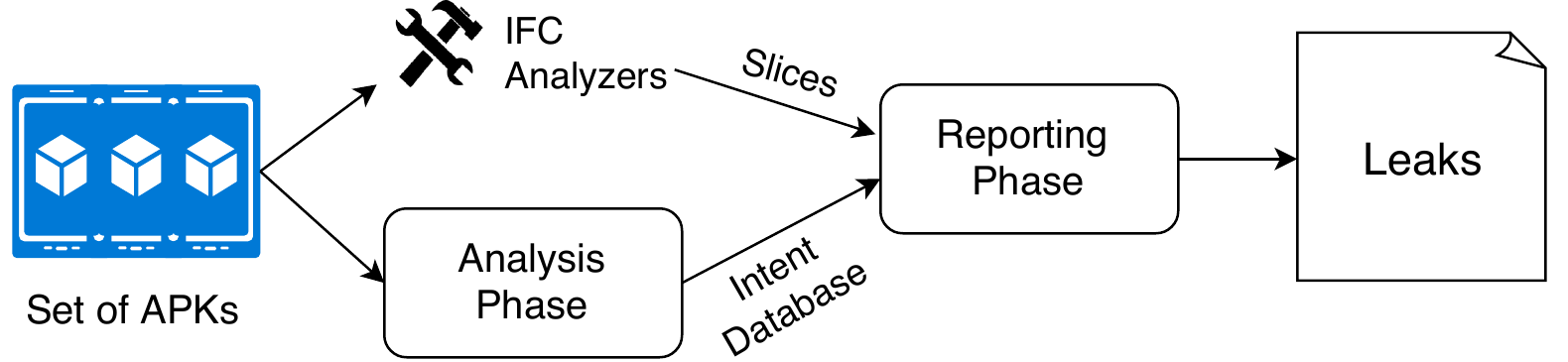}
\caption{Analysis Framework}
\label{fig:framework}
\end{figure}

The fundamental problem of intent analysis for static analysis is the dynamic nature of intents. Static IFC analyses generally leverage dataflow analyses like backwards slicing to determine whether sensitive information (e.g., a device id) may flow at a sink (e.g., internet). However, if a slice contains statements where data is extracted from a received intent, it cannot determine the data's sensitivity without detailed knowledge on possible senders and their semantics. 

Figure~\ref{fig:framework} presents the major building blocks of our analysis framework. In the \emph{analysis phase} a set of APKs under inspection (e.g., all apps installed on a device) is processed and the extracted information stored into a SQL database named \emph{IntentDB}. We collect two sets of information, app-specific information, i.e., package and class name, and registered intent filters to receive implicit intents, as well as intent sender-specific information, i.e., information required to identify potential receiver(s), key, type, and the actual data being sent.

The database is fed into the \emph{reporting phase} together with the receiving app's information flows from a baseline (intra-component) IFC analyzer. If a flow originates at a \emph{getXXXExtra} method\footnote{\emph{getXXXExtra} methods retrieve type-specific data from a received intent that has been added through the corresponding \emph{putXXXExtra} method.}
, we consider the respective sender's outbound data as the actual data source to that flow. Remember that data can only successfully be transmitted via \emph{put/getExtra} methods if the key parameters of both methods match and the signatures of the \emph{put} and \emph{get} methods correspond (e.g. the value's type of the \emph{put} method equals the return type of the \emph{get} method). 
Thus we determine all potential senders of this intent based on matching the target component or intent action. For each of these senders we extract the \emph{key}, \emph{value}, and \emph{put} signature (see section~\ref{sec:analysis}) from database. If the key and the put signature match this \emph{getXXXExtra} method invocation\footnote{The \emph{getXXXExtra}'s key is determined via backward slicing}, we determine the sensitivity of the transmitted value based on a categorization of sources. 
If the value is considered sensitive, we report a potential information flow violation. 

As an example, consider Listings~\ref{listing:intent} and~\ref{listing:intentthree} again: The \emph{sendTextMessage} (line~\ref{lst:line2_3}) receives data from the \emph{getStringExtra} method (line~\ref{lst:line2_2}). Therefore we scan our database for potential intent senders of App~B's received intent (line~\ref{lst:line2_1}): App~A sends an implicit intent with matching intent action (\emph{``CUSTOM\_INTENT.ACTION"}, line~\ref{lst:line1_3} of Listing~\ref{listing:intent}). The signature of the \emph{putExtra} method (line~\ref{lst:line1_4}) has a \emph{String} parameter, which matches the return type of the \emph{getStringExtra} method of the receiver, and the keys of the sender and receiver (``data'') match. Thus, the source of the transmitted value (i.e. the IMEI of the device) is considered the information source of the flow to the SMS transmission in App~B. As the IMEI is sensitive information, an information flow violation is reported.

\textbf{Structure of \emph{IntentDB}:} Table~\ref{table:intent_extract} shows an example entry (from the Telegram messenger app) of the database
. As apps consist of several classes, this table has potentially multiple entries for the same app. All entries belonging to one app can be identified by the unique package name. Similarly, each class can send out several intents and hence for each intent sent we will list a separate entry (package name \& class name are the same). The column ``Put Signature" is considered for mapping the \emph{put} method to the corresponding \emph{getXXXExtra} method at the time of intent resolution. 
Depending on the non-empty fields, an entry in the database represents an intent receiver and/or sender. If the \emph{Intent Filter} field is set, the app may receive intents. If either the \emph{Target Component} or the \emph{Intent Action} field is set, it acts as an intent sender.

\begin{table*}
\caption{Example database for a class that can receive as well as send intents}
\centering
\begin{adjustbox}{width=\textwidth}
	\small
\begin{tabular}{ | m{1.35cm} | m{1.65cm} | m{2.7cm} | m{1.2cm} | m{2.3cm} | m{1.2cm} | m{1.8cm} | m{2.5cm} | } 
 \hline
 \centering Package Name & \centering Class Name & \centering Intent Filter & \centering Target Component & \centering Intent Action & \centering Key & \centering Value & {\centering Put Signature} \\
 \hline
 \hline
 \scriptsize{org.telegram. messenger} & \scriptsize{Firebase In- stance\-Id Service} & \scriptsize{com.google.firebase.IN\-STANCE\_ID\_EVENT} & \scriptsize{null} & \scriptsize{com.google.android. gcm.intent.SEND} & \scriptsize{"google.to"} & \scriptsize{String url = "google.com/iid"} & \scriptsize{putExtra (String, String)} \\
\hline
\end{tabular}
\end{adjustbox}
\label{table:intent_extract}
\end{table*}

\subsection{Analysis Phase} \label{sec:analysis}
Figure~\ref{fig:workflow} depicts the workflow of the analysis phase. In the sequel, we describe the details of each component:

\begin{figure}[tb]
\includegraphics[width=\linewidth,viewport=20 30 600 145,clip]{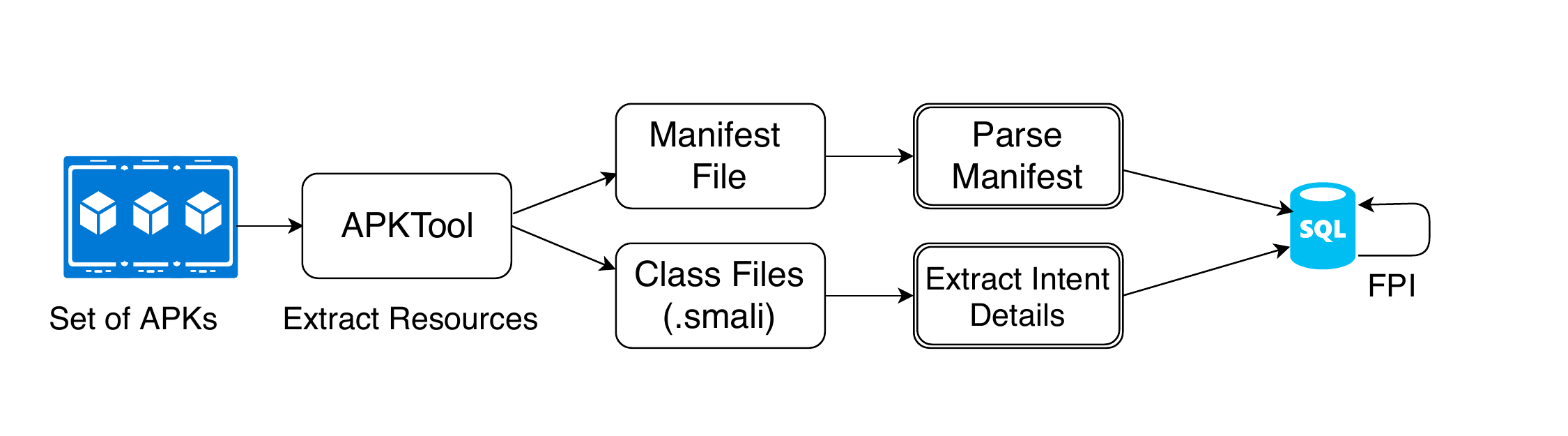}
\caption{Analysis Phase, FPI stands for fixed point iteration}
\label{fig:workflow}
\end{figure}

\subsubsection{Apktool} A set of APKs is processed by \emph{Apktool}~\cite{apktool}, which extracts and decodes the resources of an APK (e.g., \emph{manifest.xml})
. It decodes the Dalvik bytecode file (\emph{classes.dex}) of the APK to more comprehensible Smali class files~\cite{smali}. 

\subsubsection{Manifest Parser} Parsing the manifest file extracts various app details (first set of information), i.e., \emph{package and class name}, as well as \emph{supported intent filters}. This information is  mapped to the first three columns of the table and identifies potential receivers of an intent. Even though intent receivers are typically registered in the manifest file, the \emph{registerReceiver} method can register an intent receiver at runtime. In our experiment with 90 apps, we find 433 dynamically registered receivers ($\approx 5\%$ of all intent receivers). We scan class files for dynamically registered receivers and store them in IntentDB.

\subsubsection{Dynamic Intent Data Extraction} In this module, we scan each class file for methods that initiate an intent (sender methods), e.g., \emph{startActivity}. The \emph{Android documentation}~\cite{intentDocumentation} defines 25 such methods including 12 variants of \emph{startActivity}, 11 variants of \emph{broadcast}, \emph{startService} and \emph{bindService}. 

\paragraph{Identifying Target Component/Intent Action} For every sender method we compute its backward slice and trace until we find the corresponding intent initialization(s). The goal is to identify its target component (for an explicit intent) or intent action (implicit intent). The intent type depends on the intent's constructor but can be altered using the \emph{explicit-transformation} methods \emph{makeMainActivity}, \emph{makeRestartActivityTask}, \emph{setClass}, \emph{setClassName}, \emph{setComponent}, \emph{setPackage} or \emph{setSelector}, which can also \emph{change} the target component after the fact. We analyze these cases to extract the actual target:
In the case of an explicit intent, we identify the name of the target component. For an implicit intent, we extract the intent action. Any app defining this intent action as supported intent filter (dynamically or in its manifest file) is a potential receiver of this intent. Unfortunately, one cannot always statically determine intent details (e.g., intent action) as they may be influenced by runtime information, which is a general limitation of static analysis. We conservatively approximate such situations, i.e., may include several potential intent actions into the database. Future work may rule out non-matching substrings of potential target name/action strings similar to reflection analyis~\cite{grech18ECOOP}.

\paragraph{Identifying Key-Value Pairs}\label{keyvalue} There are several methods to associate extra data with an intent, generally leveraging \emph{key-value} pair schemes. Senders register a value specifying the key, e.g., \emph{Intent.putExtra(``Test-Key", ``Test-Value")} will register the string \emph{``Test-Value"} as data for the key \emph{``Test-Key"}, which can be extracted by a corresponding receiver using the \emph{Intent.getExtra(``Test-Key")} method. Trying to receive a key with a non-matching data type results in no value being transmitted. Therefore precise analysis mandates a correct matching of \emph{get} and \emph{put} methods. Unlike related work~\cite{li2015iccta,wei2014amandroid} we handle the respective \emph{put/get} method pairs for all basic data types and store the precise signature of any \emph{put} method in \emph{IntentDB} to consider matching types and keys when resolving values received by \emph{getXXXExtra} methods at intent receivers.

\begin{lstlisting}[language=Java, belowskip=-0.8 \baselineskip, caption= Sender: OutFlowActivity,label={listing:appA}, float=tb, escapeinside={@}{@}]
// APP A (OutFlowActivity)
@\label{appA:line1_1}@TelephonyManager tel = (TelephonyManager) getSystemService(TELEPHONY_SERVICE);
@\label{appA:line1_2}@String imei = tel.getDeviceId(); // source
@\label{appA:line1_3}@Intent i = new Intent("action_test");
@\label{appA:line1_4}@i.putExtra("data", imei);
@\label{appA:line1_5}@startActivity(i); // sink	

// APP B (Intermediate Activity) -- Capable of receiving "action_test"
@\label{appB:line2_1}@Intent i = getIntent();
@\label{appB:line2_2}@String imei = i.getStringExtra("data");
@\label{appB:line2_3}@Intent newIntent = new Intent("action_test2");
@\label{appB:line2_4}@newIntent.putExtra("secret", imei);
@\label{appB:line2_5}@startActivity(newIntent);

//APP C (InFlow Activity) -- Capable of receiving "action_test2"
@\label{appC:line2_1}@Intent i = getIntent();
@\label{appC:line2_2}@String imei = i.getStringExtra("secret");
@\label{appC:line2_3}@smsManager.sendTextMessage("1234567890", null, imei, null, null); // sink
\end{lstlisting} 

\subsubsection{Fixed Point Iteration} Intent communication may involve more than two apps/components. In our experiments with 90 apps, we find 54 cases where more than two components were involved in a transitive information flow. In such a case, \emph{IntentDB} contains a \emph{getXXXExtra} method in the column \emph{Value}. For example, in Listings~\ref{listing:appA}, app A is sending the device id (secret data) to app B. App B forwards this data to app C, and finally app C leaks it via an SMS. The first 3 rows of Table~\ref{caseATable} show the table \emph{IntentDB} prior to fixed point iteration. To resolve transitive flows through multiple components we iterate in a fixed point iteration through the entries of \emph{IntentDB} for which \emph{Value} contains a \emph{getXXXExtra} method. The \emph{com.appB} entry in Table~\ref{caseATable} is such an example where data from a received intent is being sent out via another intent. In order to identify the received data, we determine all apps from which this component could receive the intent on which \emph{getXXXExtra} is invoked. In our example \emph{com.appB} receives from \emph{com.appA}. Finally we match the corresponding \emph{key}-\emph{value} pair through their \emph{get-put} signatures and create a new entry, replacing the original source (\emph{getXXXExtra} method) by the transmitted value. The created entry for our example is shown in gray in Table~\ref{caseATable}. To accommodate for modular analysis and thus potential new compatible senders, we retain the old database entry (row 2). The reporting phase described in the next section now matches the added row with the intent received in App C to reveal the transitive information flow of sensitive data to the SMS sink.

\begin{table*}
\caption{\emph{IntentDB} for Listing~\ref{listing:appA}. Fixed point iteration adds the last row }
\centering
\begin{adjustbox}{width=\textwidth}
	\small
\begin{tabular}{ | l | l | l | l | l | l | l | l | } 
 \hline
 Pckg. Name & Class Name & Intent Filter & Target Component & Intent Action & Key & Value & Put Signature \\ 
 \hline
 \hline
 com.appA & OutFlow Activity & null & null & action\_test & "data" & \emph{Device ID} & {putExtra (String, String)}\\ 
 \hline
 com.appB & Interm. Activity & action\_test & null & action\_test2 & "secret" & getStringExtra("data") & {putExtra (String, String)}\\ 
 \hline
 com.appC & InFlow Activity & action\_test2 & null & null & null & null & null\\ [0.3ex] 
 \hline
 \color{gray}com.appB & \color{gray}Interm. Activity & \color{gray}action\_test & \color{gray}null & \color{gray}action\_test2 & \color{gray}"secret" & \color{gray}\emph{Device ID} & \color{gray}{putExtra (String, String)}\\ 
 \hline
\end{tabular}
\end{adjustbox}
\label{caseATable}
\end{table*}

\subsection{Reporting Phase}
In the reporting phase, we process information flows 
obtained by a baseline IFC analyzer together with the \emph{IntentDB} from the analysis phase. 
For ICC/IAC we are only interested in flows with sources that are potential intent receivers, 
i.e., a \emph{getXXXExtra} method (together with its key and signature). For every \emph{getXXXExtra} method in a reported information flow, we extract all potential senders to this receiver from \emph{IntentDB}, i.e., apps that use an intent with a matching target component or a matching intent action. Finally, we match \emph{get-put} method pairs and \emph{keys} to determine senders that actually send data to this receiver and report it as a (potential) leak if the transmitted data stems from a sensitive source\footnote{We utilize the categorization of sources and sinks from R-Droid~\cite{backes2016r}}.

For example, data flows from the \emph{getStringExtra} method of the intent received on line~\ref{appC:line2_1} to the data sink \emph{sendTextMessage} in App C. Our analysis thus matches any sender of the intent action \emph{action\_test2} and finds two rows in IntentDB (Table~\ref{caseATable}). We check whether any of those uses the key \emph{secret}, which both of them do. Then we match the signature of \emph{getStringExtra} with the sender's Put Signature, where again both match. Finally, we verify if one of the potentially transmitted values (\emph{Device ID, getStringExtra(”data”)}) is sensitive, thus reporting the former as an illicit information flow.

\begin{lstlisting}[caption= Sender: OutFlowActivity,label={listing:startActivityforresult}, float=tb, escapeinside={@}{@}]
public class OutFlowActivity extends Activity{
	protected void onCreate(Bundle savedInstanceState) { // ...
    @\label{lst:line7_1}@TelephonyManager tel = (TelephonyManager) getSystemService(TELEPHONY_SERVICE);
    @\label{lst:line7_2}@String imei = tel.getDeviceId(); // source
    @\label{lst:line7_3}@Intent i = new Intent(this, InFlowActivity.class);
    @\label{lst:line7_4}@i.putExtra("data", imei);
    @\label{lst:line7_5}@startActivityForResult(i, 1); 
	}
	@\label{lst:line7_6}@protected void onActivityResult(int requCd, int resCd, Intent data) {
		@\label{lst:line7_7}@String imei = data.getStringExtra("data");
		@\label{lst:line7_8}@smsManager.sendTextMessage("1234567890", null, imei, null, null); // sink
}}
\end{lstlisting} 
\begin{lstlisting}[caption= Receiver: InFlowActivity,label={listing:startActivityforresultrec}, float=tb, escapeinside={@}{@}]
public class InFlowActivity extends Activity {
	protected void onCreate(Bundle savedInstanceState) { // ...
    @\label{lst:line8_1}@Intent i = getIntent();
    @\label{lst:line8_2}@setResult(1, i);
    @\label{lst:line8_3}@finish();
}}
\end{lstlisting}   

\subsubsection{Handling of \emph{startActivityForResult} and \emph{bindService}}\label{sec:callback}
\emph{startActivityForResult} is a special case of intent communication illustrated via a code snippet of an activity in Listing~\ref{listing:startActivityforresult} (adapted from~\cite{droidBench}). \emph{OutFlowActivity} (line~\ref{lst:line7_3}) creates an explicit intent with \emph{InFlowActivity} as the target component. This intent is provided extra data \emph{imei} (line~\ref{lst:line7_4}), containing the actual \emph{IMEI} of the device (lines~\ref{lst:line7_1},~\ref{lst:line7_2}). \emph{startActivityForResult} triggers this intent (line~\ref{lst:line7_5}) with a second argument that is a \emph{request code} identifying this request.   
Listing~\ref{listing:startActivityforresultrec} contains the code snippet for the activity \emph{InFlowActivity},  receiving this intent (line~\ref{lst:line8_1}). The \emph{setResult} method (line~\ref{lst:line8_2}) returns the received intent with the same \emph{request code}.
Upon successful creation of \emph{InFlowActivity} control returns to the \emph{onActivityResult} (line~\ref{lst:line7_6} of listing~\ref{listing:startActivityforresult}) method of \emph{OutFlowActivity}. The third parameter (\emph{data}) of this method corresponds to the intent returned via the \emph{setResult} method of \emph{InFlowActivity}. This intent, originally sent by \emph{OutFlowActivity}, still contains the secret \emph{IMEI} of the device. The \emph{IMEI} is extracted (line~\ref{lst:line7_7}) and leaked (line~\ref{lst:line7_8}) via a text message.
Thus data flows from the sender to the receiver and back, as modeled in our information flow analysis.

Similarly, after a Service has been successfully bound via \emph{bindService}, control will return to an \emph{onServiceConnected} method (of an object designated as the second parameter of the original \emph{bindService} call.) This method is being passed an argument from the service intent receiver, and thus data can be returned to the intent sender. In order to identify such flows soundly and precisely our analysis models the data flow according to these patterns at both sides of the communication.

\subsection{Domain Knowledge for Java String Class and List Analysis}

As the ability to precisely determine intent senders and receivers depends significantly on the ability to identify the String values of target components or intent actions, we enrich IIFA with domain knowledge on the Java String class. IIFA understands the Smali signature of String methods and applies partial evaluation in order to recover strings created by concatenation, substring, and other String manipulation methods. Concretely, it extracts parameters, applies the respective functionality and returns the resulting string. More contrived examples like converting a string to an array of chars (to be manipulated) are beyond the scope of our tool as we are currently not targeting obfuscated code. Due to our modular design we could also add more expensive analyses like SMT-solvers that handle more cases. However, there are always undecidable cases like encrypted strings or dynamic input.

Similarly, we encode domain knowledge on the API of \emph{LinkedList} to be able to extract list entries from a given index they were stored in. Again, a more precise model of Lists improves analysis precision but in general this problem is undecidable. Other collections could be modeled analogously, which we are planning as future work.

\section{Evaluation}

We empirically evaluated our tool, IIFA, in two steps:

\begin{itemize}
\item \emph{Comparative evaluation on benchmark sets.} We applied IIFA to three standard evaluation sets for intent communication comprising 41 test cases with ground truth results for each test. We compared the precision and soundness of IIFA to 5 state of the art tools that support intent analysis.

\item \emph{Evaluation on real-world apps from the Google Playstore.} We applied IIFA to the 90 most popular apps from the Google Playstore in order to evaluate its scalability on real-world apps.
\end{itemize}

All experiments were performed on a MacBook Pro with a 2,9 GHz Intel Core i7 processor and 16 GB DDR3 RAM and MacOS High Sierra 10.13.1 installed. We used a version 1.8 JVM with 4 GB maximum heap size. 

\subsection{RQ1: Precision and Soundness of IIFA}

\subsubsection{Benchmark evaluation datasets}
In order to evaluate whether the precision and soundness of our approach are on par with state of the art analyses, we use three separate benchmark sets to compare the results of our tool to related approaches. However, as IIFA is not a stand-alone IFC tool but rather only models the information flows through intents, we restrict ourselves to the subset of the three benchmark sets that test the results of intent-based communication:

\begin{itemize}
\item The intent-related cases of the original DroidBench test suite~\cite{droidBench} \textbf{(14 test cases)}
\item The extension proposed by IccTA~\cite{li2015iccta} \textbf{(18 test cases)}
\item ICC-Bench, proposed by Wei et al.~\cite{wei2014amandroid} \textbf{(9 test cases)}
\end{itemize}

Note that the mentioned benchmark sets include several advanced usage scenarios of intents. An example of these scenarios is the usage of callback methods that are triggered after an event has been delivered to its target
, which requires information tracking at both sender and receiver sides (see Section~\ref{sec:callback}). Another challenge is string manipulation, e.g., of keys for intent extra data. Finally one case passes an intent with sensitive data through multiple components before finally leaking the stored data. The authors of each benchmark set provide ground truth for each test case, which we use to measure precision and soundness. 

\subsubsection{Comparative evaluation}
Based on true positives ($\mathit{tp}$), false positives ($\mathit{fp}$), and false negatives ($\mathit{fn}$) we use the following metrics to compare the performance of IIFA with the related tools:
\[
\begin{array}{ccc}
 \text {\bf Precision} & \text{\bf Recall} & \mathit{F_1}\text{\bf-measure} \\
 p = \frac{\mathit{tp}}{\mathit{tp} + \mathit{fp}} & r = \frac{\mathit{tp}}{\mathit{tp} + \mathit{fn}} & \frac{2 p r}{p + r}   
\end{array}
\]

We applied IIFA to the original DroidBench benchmark set, where 14 test cases are relevant for intent communication. On these benchmarks IIFA achieved precision and recall ratios of 100\%. Given that the DroidBench tests cover almost identical cases as those in IccTA and ICC-Bench, we do not list details on the results for space reasons.
We further applied IIFA to the IccTA extension of Droidbench~\cite{droidBench} and ICC-Bench~\cite{wei2014amandroid}, and compared the results to the five most prominent tools for Android intent information flow analysis: FlowDroid~\cite{arzt2014flowdroid}, AppScan~\cite{appscan} and IccTA~\cite{li2015iccta} are limited to ICC, 
DidFail~\cite{klieber2014android} and Amandroid~\cite{wei2014amandroid} come with their own inter-app analysis. 
Table~\ref{table:1} displays the results of the different tools on both benchmark sets (related work taken from Li et al.~\cite{li2015iccta}), which also summarizes the metrics. 
In sequel, we compare the results of Amandroid, IccTa and IIFA (Table~\ref{table:1}).
%
%
%

\textbf{Amandroid} Amandroid has a mediocre recall rate of 60\% due to imprecision in complex sender functions (section~\ref{sec:callback}), imprecision in the lifecycle model of Services (\emph{startService2}) and string manipulation (\emph{DynRegister2}) and lacking support for Content Providers. Its IAC capabilities are not tested in any benchmark.

\textbf{IccTA} IccTA in general has good precision and recall ratios, but fails on test cases that include complex string manipulations, e.g., 
\emph{DynRegister2} and 
\emph{startActivity7}
. 

In our own experiments IccTA and Amandroid failed to resolve the key and/or type of ICC (see section~\ref{keyvalue}), confirming the fact that this is not mentioned in their publications. 

\textbf{Our approach: IIFA} 
To evaluate the precision and soundness of IIFA we applied it to the benchmark sets. We used R-Droid~\cite{backes2016r}, a static IFC tool, to generate the intra-app flows, and compared our analysis results with the provided ground truth. We also verified the analysis results manually. As expected it also resolved our own tests with keys and types correctly.

\noindent\framebox{\parbox{\dimexpr\linewidth-2\fboxsep-2\fboxrule}{%
\textbf{Answer to RQ1:}\itshape  \,Our tool IIFA is on par with (or even outperforms) all related state-of-the-art tools on the first three benchmark sets, and resolves our own experiments while the most precise related works suffer from lacking type/key matching.}}

Clearly these benchmarks, 
gathered from related work, do not involve typically ignored (as hard to analyze) language features like reflection, or native code. However, they do cover a range of difficult features with respect to intent analysis, like dynamically composed strings, arrays with both sensitive and insensitive data, etc.
The intention of this research question was to assert that the scalability gains (see RQ2) do not negatively impact other essential properties of our analysis, and to demonstrate the impact of precise matching of types and keys for transmitted intent data. Even though small, this microbenchmark evaluation demonstrates that we are more than on par with the best related work.

\begin{table}
\caption{Comparison results.  \textcolor{green}{\checkmark}: true positive, \textcolor{red}{$\star$}: false positive, \textcolor{red}{$\circ$}: false negative }
\centering
 \begin{tabular}{| m{3.4cm} |  m{1cm} | m{1.1cm} | m{1.1cm} | m{0.6cm} | m{0.8cm} | m{.7cm} | m{0.6cm}|} 
 \hline
 \multicolumn{8}{|c|}{IccTA} \\
 \hline
 Test Case &  Explicit ICC & Flow Droid & App Scan & Did Fail & Aman droid & IccTA & Our Tool \\ [0.5ex] 
 \hline\hline
 
 startActivity1  & T & \textcolor{green}{\checkmark} \textcolor{red}{$\star$} & \textcolor{green}{\checkmark} \textcolor{red}{$\star$} & \textcolor{red}{$\circ$} & \textcolor{green}{\checkmark} & \textcolor{green}{\checkmark} & \textcolor{green}{\checkmark}\\ 
 \hline
 
 startActivity2  & T & \textcolor{green}{\checkmark} (4 \textcolor{red}{$\star$}) & \textcolor{green}{\checkmark} (4 \textcolor{red}{$\star$}) & \textcolor{red}{$\circ$} & \textcolor{green}{\checkmark} & \textcolor{green}{\checkmark} & \textcolor{green}{\checkmark}\\
 \hline
 
 startActivity3  & T & \textcolor{green}{\checkmark}(32 \textcolor{red}{$\star$}) & \textcolor{green}{\checkmark}(32 \textcolor{red}{$\star$}) & \textcolor{red}{$\circ$} & \textcolor{green}{\checkmark} & \textcolor{green}{\checkmark} & \textcolor{green}{\checkmark}\\
 \hline
 
 startActivity4  & F & \textcolor{red}{$\star$} \textcolor{red}{$\star$} & \textcolor{red}{$\star$} \textcolor{red}{$\star$} & \textcolor{red}{$\star$} &  &  & \\
 \hline
 
 startActivity5   & F & \textcolor{red}{$\star$} \textcolor{red}{$\star$} & \textcolor{red}{$\star$} \textcolor{red}{$\star$} & \textcolor{red}{$\star$} &  &  & \\
 \hline
 
 startActivity6   & T & \textcolor{red}{$\star$} \textcolor{red}{$\star$} & \textcolor{red}{$\star$} \textcolor{red}{$\star$} &  & \textcolor{red}{$\star$} &  & \\
 \hline
 
 startActivity7   & T & \textcolor{red}{$\star$} \textcolor{red}{$\star$} & \textcolor{red}{$\star$} \textcolor{red}{$\star$} &  & \textcolor{red}{$\star$} & \textcolor{red}{$\star$} & \\
 \hline
 
 startActivityForResult1   & T & \textcolor{green}{\checkmark} & \textcolor{green}{\checkmark} & \textcolor{red}{$\circ$} & \textcolor{green}{\checkmark} & \textcolor{green}{\checkmark} & \textcolor{green}{\checkmark}\\
 \hline
 
 startActivityForResult2   & T & \textcolor{green}{\checkmark} & \textcolor{red}{$\circ$} & \textcolor{red}{$\circ$} & \textcolor{red}{$\circ$} & \textcolor{green}{\checkmark} & \textcolor{green}{\checkmark}\\
 \hline
 
 startActivityForResult3   & T & \textcolor{green}{\checkmark} \textcolor{red}{$\star$} & \textcolor{red}{$\circ$} & \textcolor{red}{$\circ$} & \textcolor{red}{$\circ$} \textcolor{red}{$\star$}  & \textcolor{green}{\checkmark} & \textcolor{green}{\checkmark}\\ [1ex] 
 \hline
 
 startActivityForResult4   & T & \textcolor{green}{\checkmark} \textcolor{green}{\checkmark} \textcolor{red}{$\star$} & \textcolor{green}{\checkmark} \textcolor{red}{$\circ$} & \textcolor{red}{$\circ$} \textcolor{red}{$\circ$} & \textcolor{green}{\checkmark} \textcolor{red}{$\circ$} \textcolor{red}{$\star$} & \textcolor{green}{\checkmark} \textcolor{green}{\checkmark} & \textcolor{green}{\checkmark} \textcolor{green}{\checkmark}\\ 
 \hline
 
 startService1   & T & \textcolor{green}{\checkmark} \textcolor{red}{$\star$} & \textcolor{green}{\checkmark} \textcolor{red}{$\star$} & \textcolor{red}{$\circ$} & \textcolor{green}{\checkmark} & \textcolor{green}{\checkmark} & \textcolor{green}{\checkmark}\\
 \hline
 
 startService2   & T & \textcolor{green}{\checkmark} \textcolor{red}{$\star$} & \textcolor{green}{\checkmark} \textcolor{red}{$\star$} & \textcolor{red}{$\circ$} & \textcolor{red}{$\circ$} & \textcolor{green}{\checkmark} & \textcolor{green}{\checkmark}\\
 \hline
 
 bindService1   & T & \textcolor{green}{\checkmark} \textcolor{red}{$\star$} & \textcolor{green}{\checkmark} \textcolor{red}{$\star$} & \textcolor{red}{$\circ$} & \textcolor{red}{$\circ$} & \textcolor{green}{\checkmark} & \textcolor{green}{\checkmark}\\
 \hline
 
 bindService2  & T & \textcolor{red}{$\circ$} & \textcolor{red}{$\circ$} & \textcolor{red}{$\circ$} & \textcolor{red}{$\circ$} & \textcolor{green}{\checkmark} & \textcolor{green}{\checkmark}\\ [1ex] 
 \hline
 
 bindService3  & T & \textcolor{red}{$\circ$} & \textcolor{red}{$\circ$} & \textcolor{red}{$\circ$} & \textcolor{red}{$\circ$} & \textcolor{green}{\checkmark} & \textcolor{green}{\checkmark}\\
 \hline
 
 bindService4  & T & \textcolor{green}{\checkmark} \textcolor{red}{$\star$} \textcolor{red}{$\circ$} & \textcolor{green}{\checkmark} \textcolor{red}{$\star$} \textcolor{red}{$\circ$} & \textcolor{red}{$\circ$} \textcolor{red}{$\circ$} & \textcolor{red}{$\circ$} \textcolor{red}{$\circ$} & \textcolor{green}{\checkmark} \textcolor{green}{\checkmark} & \textcolor{green}{\checkmark} \textcolor{green}{\checkmark}\\
 \hline
 
 sendBroadcast1 & F & \textcolor{green}{\checkmark} \textcolor{red}{$\star$} & \textcolor{green}{\checkmark} \textcolor{red}{$\star$} & \textcolor{red}{$\circ$} & \textcolor{green}{\checkmark} & \textcolor{green}{\checkmark} & \textcolor{green}{\checkmark}\\
 \hline
 
 \multicolumn{8}{|c|}{ICC-Bench} \\
 \hline
 
 Explicit1  & T & \textcolor{green}{\checkmark} & - & \textcolor{red}{$\circ$} & \textcolor{green}{\checkmark}  & \textcolor{green}{\checkmark} & \textcolor{green}{\checkmark}\\ [1ex] 
 \hline 
 
 Implicit1  & F & \textcolor{green}{\checkmark} & - & \textcolor{green}{\checkmark} & \textcolor{green}{\checkmark}  & \textcolor{green}{\checkmark} & \textcolor{green}{\checkmark}\\ [1ex] 
 \hline
 
 Implicit2  & F & \textcolor{green}{\checkmark} & - & \textcolor{green}{\checkmark} & \textcolor{green}{\checkmark}  & \textcolor{green}{\checkmark} & \textcolor{green}{\checkmark}\\ [1ex] 
 \hline
 
 Implicit3  & F & \textcolor{green}{\checkmark} & - & \textcolor{green}{\checkmark} & \textcolor{green}{\checkmark}  & \textcolor{green}{\checkmark} & \textcolor{green}{\checkmark}\\ [1ex] 
 \hline
 
 Implicit4  & F & \textcolor{green}{\checkmark} & - & \textcolor{green}{\checkmark} & \textcolor{green}{\checkmark}  & \textcolor{green}{\checkmark} & \textcolor{green}{\checkmark}\\ [1ex] 
 \hline
 
 Implicit5  & F & \textcolor{green}{\checkmark} \textcolor{red}{$\star$} & - & \textcolor{green}{\checkmark} & \textcolor{green}{\checkmark}  & \textcolor{green}{\checkmark} & \textcolor{green}{\checkmark}\\ [1ex] 
 \hline
 
 Implicit6  & F & \textcolor{green}{\checkmark} & - & \textcolor{green}{\checkmark} & \textcolor{green}{\checkmark}  & \textcolor{green}{\checkmark} & \textcolor{green}{\checkmark}\\  [1ex] 
 \hline
 
 DynRegister1  & F & \textcolor{red}{$\circ$} & - & \textcolor{red}{$\circ$} & \textcolor{green}{\checkmark}  & \textcolor{green}{\checkmark} & \textcolor{green}{\checkmark}\\  [1ex] 
 \hline
 
 DynRegister2 & F & \textcolor{red}{$\circ$} & - & \textcolor{red}{$\circ$} & \textcolor{red}{$\circ$}  & \textcolor{red}{$\circ$} & \textcolor{green}{\checkmark}\\ [1ex] 
 \hline
 
 \multicolumn{8}{|c|}{Sum, Precision, Recall and F1} \\
 \hline
 
 \multicolumn{2}{| l |}{\textcolor{green}{\checkmark}, higher is better}   & 20 & 10 & 6 & 15 & 24 & 25\\
 \hline
 
 \multicolumn{2}{| l |}{\textcolor{red}{$\star$}, lower is better }  & 53 & 50 & 2 & 4 & 1 & 0\\
 \hline
 
 \multicolumn{2}{| l |}{ \textcolor{red}{$\circ$}, lower is better } & 5 & 6 & 19 & 10 & 1 & 0\\
 \hline
 
 \multicolumn{2}{| l |}{ Precision $p = \textcolor{green}{\checkmark} / (\textcolor{green}{\checkmark} + \textcolor{red}{\star})$ }  & 27.3\% & 16.7\% & 75\% & 78.9\% & 96\% & 100\%\\
 \hline
 
 \multicolumn{2}{| l |}{ Recall $r = \textcolor{green}{\checkmark} / (\textcolor{green}{\checkmark} + \textcolor{red}{\circ})$   } & 80\% & 62.5\% & 24\% & 60\% & 96\% & 100\%\\
 \hline
 
 \multicolumn{2}{| l |}{ F$_1$-measure $2pr/(p + r)$   } & 0.41 & 0.26 & 0.36 & 0.68 & 0.96 & 1\\
 \hline
 
\end{tabular}
\label{table:1}
\end{table}
\subsection{RQ2: Evaluating the scalability of IIFA}

\subsubsection{Evaluation on real-world apps}
\label{subsec:real-world-apps}
\begin{figure}[tb]
\includegraphics[width=\linewidth]{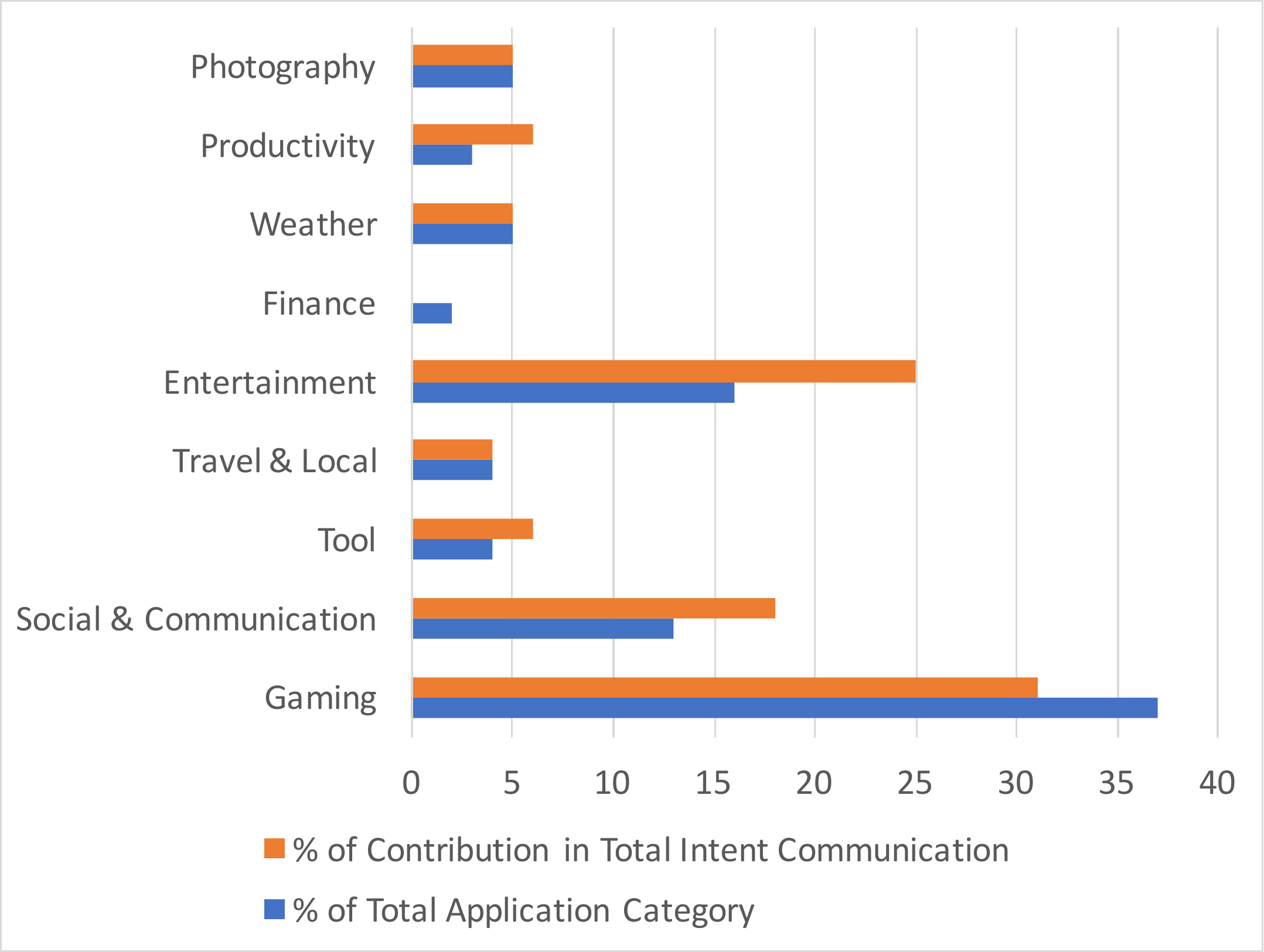}
\caption{Intent Usage}
\label{fig:stats}
\end{figure}

We applied IIFA to real-world apps to assess whether the analysis scales to a realistic corpus of large real-world apps.
We therefore downloaded the 90 most popular apps from the Google playstore. IIFA successfully analyzes each of these apps.
We compared the scalability of IIFA with the most related approach IccTA~\cite{li2015iccta} on real world apps. We chose IccTA as it achieves significantly better precision and soundness numbers in the benchmark evaluation, as Amandroid's median runtime and IccTA's are similar~\cite{li2015iccta}, and
as most other tools do not support ICC analysis for more than two components
. To analyze inter-app communication, IccTA's GitHub page proposes \emph{ApkCombiner}~\cite{apkcombiner}, which merges two or more apps into one. However, in our experiments \emph{ApkCombiner} only worked well with small numbers of benchmark set apps. When applied to numerous pairs of real world apps, it 
failed with an error message. But even if the joining process was not problematic, it would aggravate the single-app scalability issues~\cite{li2015iccta}, which were confirmed in a separate recent comparative study~\cite{qiu2018analyzing}, and would therefore lead to combinatorial explosion of apps to be analyzed. In our study with the top 90 apps from \emph{Google Play Store}, we find that 60\% of all intents are implicit. IccTA would have to eagerly merge all combinations of (at least) two complete apps. This is at least 8,100 combinations for our 90 apps already, most of which are not communicating.

In contrast, IIFA analyzes the communication compositionally based on small summaries in a database and combines only the transmitted ICC/IAC data with the intra-app flows of respective intent receivers.
The average per app execution time of IIFA over 90 apps is 87.91 seconds. On average, the analysis phase took 46.81 seconds (maximum 52.20, minimum 32.40 seconds) and the reporting phase 41.10 seconds (maximum 48.60, minimum 35.10 seconds). Considering that (intra-app) static IFC analyses usually require a large amount of time to analyze real world apps IIFA's additional cost is quite feasible for a realistic usage scenario. In our experiments with related tools such as IccTA, R-Droid, and AmanDroid, information flow analysis of a real-world app from our test set consistently took more than 30 minutes. Let us assume \emph{\tIFC} is the total time taken by an IFC analyzer to analyze one app, which according to our experiments is at least 30 minutes. IIFA adds approximately 1.5 minute per app. We refer to this sum of \tIFC and the time taken by IIFA as $t_\mathit{sum}$. Thus, the total time taken by IIFA for 90 apps is $90\times t_\mathit{sum}$ ($\approx$ 2 days). Whereas, IccTA takes $90\times 90 \times \tIFC$ ($\approx$ half a year) to analyze ICC between any two apps, let alone tuples of greater size. Clearly, the total time is dominated by the combinatorial explosion of merging apps. Therefore, IIFA would take at most 1.2\% of the time taken by IccTA.

\noindent\framebox{\parbox{\dimexpr\linewidth-2\fboxsep-2\fboxrule}{%
\textbf{Answer to RQ2:}\itshape  \, IIFA analysis avoids combinatorial explosion and, in an analysis of the 90 most downloaded real-world apps, is thus orders of magnitude faster than combining the bytecode programs eagerly.}}

\subsection{RQ3: Evaluating IAC in real-world apps}
In total we identified 10,669 calls to start an intent (either as an Activity, Broadcast or Service), 76\% of these leverage \emph{startActivity/startActivityForResult}. Further, the data in IntentDB show that (statically) 60\% of the intents are implicit. The Android documentation defines 42 different types of \emph{getXXXExtra} methods. IIFA determines that 54.6\% of these methods retrieve String values, either via \emph{Bundle.getString(String)} (38.8\%) or \emph{String.getStringExtra(String)} (15.7\%). For 2\% of the sent intents IIFA was unable resolve the target component or intent action and conservatively approximated it to either including multiple actions (0.6\%) or even a dummy action (1.4\%), which requires manual inspection to resolve the potential strings (e.g. due to dynamic input from a file).

Figure~\ref{fig:stats} provides a categorization of the 90 apps along with their intent usage. 37\% of the apps are \emph{Games}, which contribute the most to intent communications (31\%). Interestingly, we find that 6\% of the total intent communication is triggered by a single a \emph{Communication} app, \emph{whatsApp}. It contributes to 35\% of the intent communication in the \emph{Communication} category.

\begin{figure}[hbt!]
\includegraphics[width=.8\linewidth,viewport=55 85 722 525,clip]{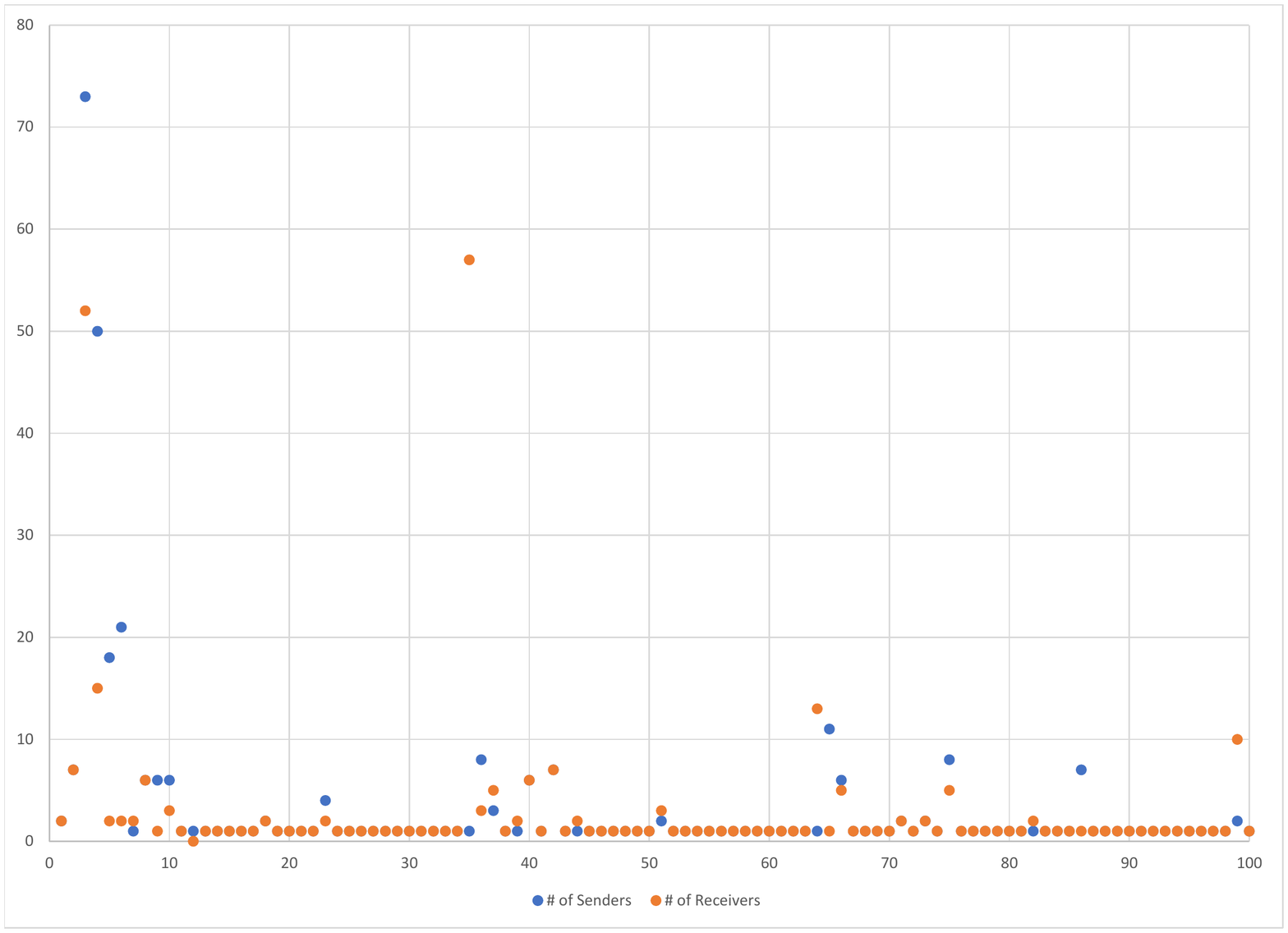}
\includegraphics[width=.8\linewidth,viewport=5 35 410 302,clip]{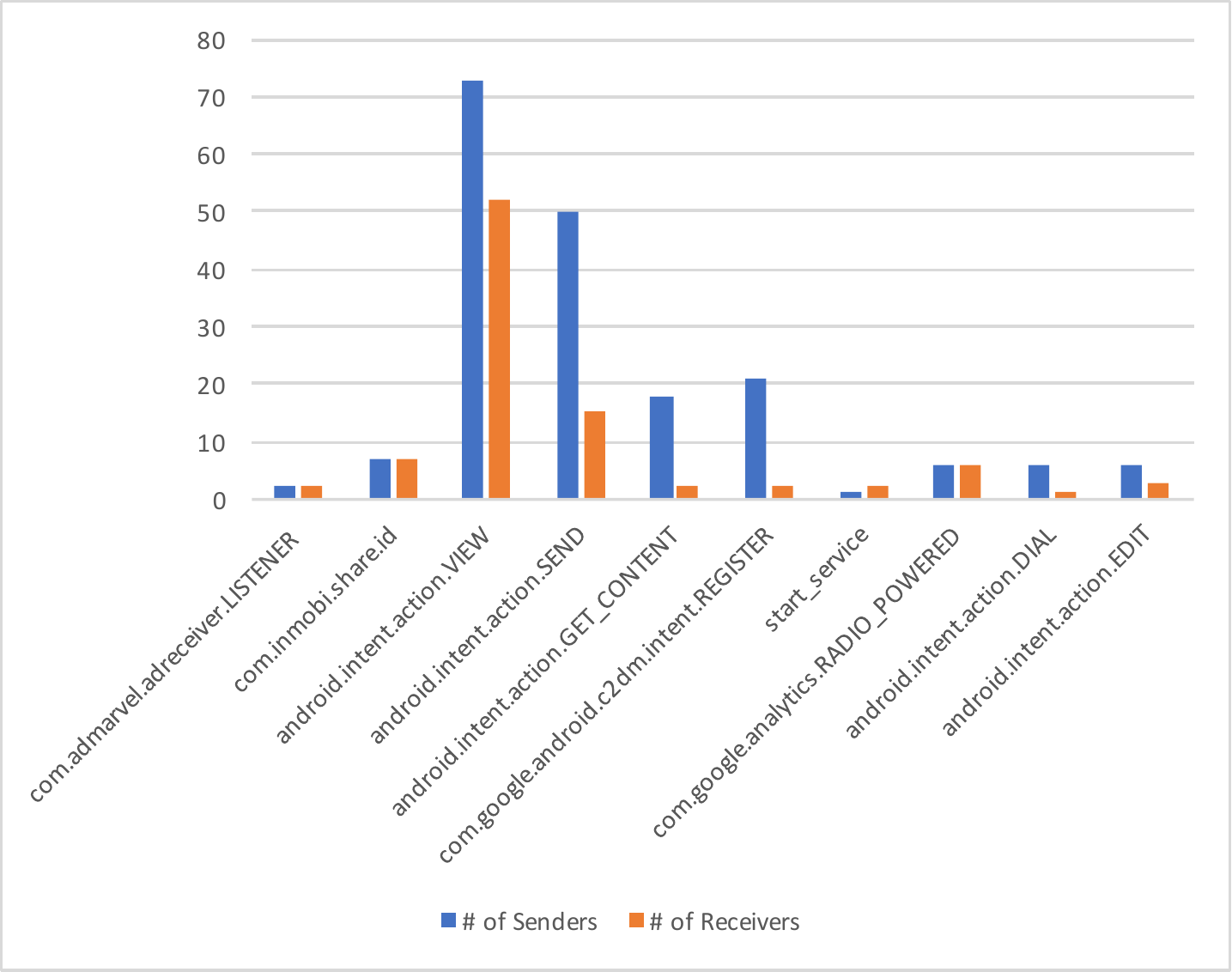}
\caption{Intent Actions (x-axis) per sender (blue) and receiver (orange, y-axis)}
\label{fig:morestats}
\end{figure}

IIFA's database contains senders with 380 distinct Intent actions, 100 out of them with corresponding receivers in our test set (excluding system apps and OS). Figure~\ref{fig:morestats} displays these 100 intent actions and their numbers of senders and receivers. The second graph zooms in on the ten most widely used intent actions. 
Note that the numbers of \emph{actual} intent receivers is lower due to type and key matching of intent extra data, demonstrating the precision of our analysis.
\emph{android.intent.action.VIEW} is most widely used (73 senders and 52 receivers) followed by \emph{android.intent.action.SEND} (50 senders and 15 receivers). \emph{android.intent.action.DIAL} is an Intent action to invoke the OS phone dialer. As \emph{Viber}, a \emph{voip} app, also registers to receives it, users will be asked to select how to make a phone call. Other apps not providing \emph{voip} ought not receive it. IIFA determines potential rogue apps via a simple database lookup. To the best of our knowledge, we are the first to predict and analyze these communication patterns.

IIFA detects 62 ICC-based information flows from sensitive sources among the 90 apps. Our results shows that even widely used apps share sensitive information via implicit intents, which may lead to \emph{intent interception} and \emph{intent hijacking} attacks~\cite{lu2012chex}. We manually validated these claims in the following apps: \emph{Katwarn} 
provides hazard and disaster warnings and has been downloaded over one million times. IIFA finds that one of it's activities (\emph{GuardianAngelService}) shares the last known location via an implicit intent. An app that registers the respective intent filter can try to intercept this intent (\emph{intent interception}) to obtain the location details without having permission to access the device's location. Similarly, the shopping app \emph{ebay} (via activity \emph{EventItemsFragment}) and the location \& travel app \emph{Google Earth} share internal device resources 
via an implicit intent, which are thus also prone to intent interception. IIFA also determines that \emph{ebay} shares sensitive \emph{tags} via both implicit (in \emph{CheckoutActivity} activity) and explicit intents (in \emph{PaypalCreditPromotionsActivity} activity). While \emph{tags} do not always contain sensitive information, they may contain sensitive objects. For example, a \emph{tag} can contain a \emph{url}, e.g., BankA.com, to be opened by an intent created by the receiver. If this \emph{tag} is shared via an implicit intent, a malicious app that registers the respective intent filter may manipulate this \emph{tag} to open another url, e.g., BankB.com (\emph{intent hijacking} as part of a phishing attack). 

\noindent\framebox{\parbox{\dimexpr\linewidth-2\fboxsep-2\fboxrule}{%
\textbf{Answer to RQ3:}\itshape  \, IIFA analyzes the IAC patterns and may detect rogue apps registering for Intent actions they are not supposed to handle. We detect a number of problematic information flows that may be abused by malevolent apps that have not been reported previously.}}

\subsection{Evaluation Summary and Discussion}
We empirically evaluated IIFA in two steps. 
IIFA does not suffer from combinatorial explosion (RQ2) as it anlalyzes each app once (with potentially one additional intra-app IFC analysis by an external tool), and its precision and recall are at least on par with related work.

Due to the nature of our analysis we rely on program slices generated from a baseline IFC analyzer. Therefore our combined analysis as presented in the evaluation section inherits all advantages and disadvantages as well as potential implementation bugs of the underlying analysis. In order to rule out any potential interference of our analysis with the baseline analysis we chose R-Droid as our baseline analysis as it is relatively precise but does not attempt to resolve intent communication on its own. At the same time as hardly any analysis tool considers the grand obstacles for static program analysis like native code and reflection, our combined analysis also lacks these features. Further, we concentrate on intent-based communication in this work and ignore other, more atypical forms of inter-component communication such as static (global) variables or content providers. Some baseline analyses support those, in which case a combination of IIFA with such a tool would also do. 

\section{Related Work}

Arzt et al.\ proposed Flowdroid~\cite{arzt2014flowdroid}, a static taint analysis tool that includes an extensive component lifecycle model. Flowdroid was originally designed for intra-component analysis and cannot analyze string manipulations. 

R-Droid~\cite{backes2016r} is an information flow analysis tool supporting multi-threading and AsyncTasks and resolves common string manipulations for information flow purposes. R-Droid does not support intents but conservatively reports every flow to an intent sender function as a leak.

AppScan~\cite{appscan} is a commercial tool to detect vulnerabilities in mobile and web apps, including information leaks in Android apps. However, it only supports intra-app ICC analysis. Further, AppScan requires the source code of the inspected apps.

IccTA~\cite{li2015iccta} leverages static taint analysis to analyze ICC flow leaks. It ignores some ``rarely used ICC methods such as startActivities''~\cite{li2015iccta}, multi-threading and slightly involved string analysis, which may lead to missed information leaks. While the IccTA paper reports no experience with IAC, their GitHub page proposes the usage of APKCombiner~\cite{apkcombiner}, but as IccTA already reports scalability issues, merging apps will aggravate this situation and may require eager combinations of all tuples of apps, resulting in combinatorial explosion.

Wei et al.\ proposed Amandroid~\cite{wei2014amandroid}, which computes control and data flow graphs to resolve intents and inlines the invoked component's lifecycle. However, Amandroid ignores several sink functions, and like IccTA, types and keys for intent data resolution.

DroidSafe~\cite{gordon2015information} improves intent resolution via precise points-to analysis and string resolution. The authors claim an extension for IAC 
but only hit at the implementation strategy and no experiments with IAC are reported. As their ICC resolution does not take types and keys into account results are imprecise.

Klieber et al.\ proposed DidFail~\cite{klieber2014android}, that analyzes information flow in Android applications. However, DidFail is limited to the analysis of Activities and implicit intents.

Zhang et al.~\cite{AndroidLeaker} propose a hybrid approach to detect intent-based privacy leaks on Android. They leverage dynamic analysis to detect ICC, which may thus miss ICCs based on coverage. They require instrumenting all apps under inspection, which may not be feasible due to self-integrity checks. Unlike IIFA, AndroidLeaker requires manual adaption of sources and sinks for each new Android version.


Epicc~\cite{octeau2013effective} analyzes Android ICC precisely but focuses on ICC-related vulnerabilities. It does not fully resolve information flows~\cite{wei2014amandroid}.
Jiang and Xuxian~\cite{jiang2013detecting} proposed ContentScope, which detects integrity and confidentiality vulnerabilities based on dataflow analysis, but is limited to Content Providers.
Li et al.~\cite{li2014automatically} proposed PCLeaks, a data-flow analysis detecting information leaks and component hijacking in Android applications. PCLeaks cannot handle several ICC sink methods such as \emph{startActivities} and multi-threaded programs.


Hunang et al.~\cite{huang2016detecting} proposed a type system to prevent data disclosure. Unlike our approach, they require annotations in the source code, which may be a burden to app developers.
Barros et al.~\cite{barros2015static} developed a type system to precisely resolve intents and reflections, which requires annotated source code. They inherit the imperfections of their underlying framework Epicc~\cite{octeau2013effective}.
ScanDroid~\cite{fuchs2009scandroid} analyzes Android intents via a constraint system. However, the lack of distinction between component contexts leads to imprecise analysis results.


DroidChecker~\cite{chan2012droidchecker} is a taint analysis tool for Android apps supporting Intents. Due to imprecise permission handling, DroidChecker is neither sound nor complete. Further it cannot handle dynamic features of Java, such as polymorphism.
Enck et al.~\cite{enck2014taintdroid} proposed TaintDroid, a dynamic taint-analysis tool that supports intents. As a dynamic approach, finding leaks of sensitive data in vulnerable apps depends on coverage. TaintDroid cannot differentiate between different sources of sensitive data.


Octeau et al.~\cite{octeau2015composite} proposed IC3, an analysis tool for Android intents, which requires a specification of the intended program behavior written in a declarative language COAL. However, writing a COAL specification requires source code access and expert knowledge.
Liu et al.~\cite{liu2017mr} proposed MR-Droid, which generates an information flow graph and computes risk scores for various vulnerabilities. However, it does not natively support groups of more than two applications and misses various details of creating ICC links.
DroidDisintegrator by Schuster et al.~\cite{schuster2016droiddisintegrator} applied dynamic analysis using a device emulator. This emulator monitors ICC, generates policies and enforces them directly on the app. However, it is limited to intra-application information flow and is prone to false positives and negatives.
Lee et al. \cite{lee2017sealant} proposed SEALANT, a hybrid approach for information flow control that aims to intercept unwanted messages at runtime. It only supports one level of taint and thus is not able to distinguish between various sources sensitive information. 
Shen et al. \cite{shen2014information} proposed an approach for information flow control that requires modifications to the operating system, adding additional flow permissions.
	
\section{Conclusion}

In this work we propose a novel approach to analyze information flows through Android's intents.
We create a database of precise intent communication summaries and match values transmitted to the receivers based on their intra-app flows. Thus, IIFA avoids the combinatorial explosion of merging all potential communication partners. We implemented our approach in a tool called IIFA and compared it to five related tools on two standard benchmark sets. IIFA's precision and recall rates are on par or even better than previous tools, which demonstrates that our scalability improvements do not come at the cost of other essential analysis properties. Finally we applied IIFA to the 90 most popular apps from the Google Playstore and showed that due to our compositional approach the runtime performance of IIFA is moderate, scales to large real-world apps, and detects precise communication patterns and illicit IAC flows.

\bibliography{Paper}

\end{document}